# Fractal Modeling and Fractal Dimension Description of Urban Morphology

Yanguang Chen

(Department of Geography, College of Urban and Environmental Sciences, Peking University, Beijing 100871, P. R. China. E-mail: chenyg@pku.edu.cn)

**Abstract**: The conventional mathematical methods are based on characteristic length, while urban form has no characteristic length in many aspects. Urban area is a measure of scale dependence, which indicates the scale-free distribution of urban patterns. Thus, the urban description based on characteristic lengths should be replaced by urban characterization based on scaling. Fractal geometry is one powerful tool for scaling analysis of cities. Fractal parameters can be defined by entropy and correlation functions. However, how to understand city fractals is still a pending question. By means of logic deduction and ideas from fractal theory, this paper is devoted to discussing fractals and fractal dimensions of urban landscape. The main points of this work are as follows. First, urban form can be treated as pre-fractals rather than real fractals, and fractal properties of cities are only valid within certain scaling ranges. Second, the topological dimension of city fractals based on urban area is 0, thus the minimum fractal dimension value of fractal cities is equal to or greater than 0. Third, fractal dimension of urban form is used to substitute urban area, and it is better to define city fractals in a 2-dimensional embedding space, thus the maximum fractal dimension value of urban form is 2. A conclusion can be reached that urban form can be explored as fractals within certain ranges of scales and fractal geometry can be applied to the spatial analysis of the scale-free aspects of urban morphology.

## 1. Introduction

Scientific research starts from description of a phenomenon, and then focuses on understanding its work principle. The simple description is based on measurements, while the complex description relies heavily on mathematical methods (Henry, 2002). In order to describe a city, we try to express it using data. Mathematical description depends measurement description, as measurement can be treated as the basic link between mathematics and empirical studies (Taylor, 1983). In order to show the results from a measurement, we should to find the characteristic scale of a thing. A characteristic scale is a special 1-dimensional measure and can be termed *characteristic length*, which can integrate a great number of numbers into a simple number. Unfortunately, in many cases, it is impossible to find a characteristic length to describe a complex system such as a city and a system of cities. If so, we should substitute scaling concept for the characteristic scale concept. Fractal geometry can be regarded as one of the best mathematical tools for scaling analysis at present.

What is a fractal? This is not a problem for many scientists who are familiar with fractals. A fractal is regarded as a shape that is made of parts similar to the whole in some way (Feder, 1988). Quantitatively, a fractal is defined as a set for which the Hausdorff-Besicovitch dimension is strictly greater than the topological dimension (Mandelbrot, 1982). These definitions are suitable for the classical fractals, which belong to what is called thin fractals. The general concept of fractals is well known, but how to understand fractals is still a problem for specific subjects such as urban geography. A fractal has no characteristic scale and cannot be described with traditional measures such as length, area, volume, and density. The basic parameter used for fractal description is fractal dimension. Because the length of coastline cannot be effectively measured, Mandelbrot (1967) put forward the concept of fractal dimension. Fractal dimension can be defined on the base of *entropy* and *correlation function* (Chen, 2019; Feder, 1988; Mandelbrot, 1982). It is actually the invariant quantity in scaling transform, and thus a parameter indicating symmetry. Where there is an immeasurable quantity, there is symmetry (Lee, 1988). The discovery of fractals is essentially a discovery of scaling symmetry, namely, the invariance under contraction or dilation transformation (Mandelbrot and Blumen, 1989). The immeasurability of the length of coastline enlightened Mandelbrot (1989) to think about the problem of contraction-dilation symmetry (Mandelbrot, 1967).

Cities and networks of cities are complex systems bearing the property of scaling symmetry. In urban studies, it is impossible to determine the length of urban boundary and the area within the urban boundary objectively (Batty and Longley, 1994; Frankhauser, 1994). In this case, it is impossible to quantify the population size of a city. The precondition of determining urban population size is to determine urban boundary line effectively. Population is one of the central variables in the study of spatial dynamics of city development (Dendrinos, 1992), and it represents the first dynamics of urban evolution (Arbesman, 2012). If we cannot measure urban population size, how can we describe a city and measure levels of urbanization? If we cannot describe a city and quantify urbanization levels, how can we understand the mechanisms of urban evolution? Fortunately, today, we can employ fractal dimension of urban form to replace urban area and urban population size. However, a new problem have emerged, that is, how to define a city fractal and determine its fractal dimension? Although fractal cities have been studied for more than 30 years, some basic problems still puzzle many theoretical geographers. This paper is devoted to answering these questions in terms of author's experience of long-term studies on fractal cities.

## 2. Fractal cities and city fractals

### 2.1 Are cities fractals?

Is the coast of Britain a real fractal line? In fact, we cannot find any real fractals (based on fractal geometry) in the real world. This is like that we cannot find circles and triangles (based on Euclidean geometry) in the real world. All of the fractal images we encounter in books and articles represent pre-fractals rather than real fractals in mathematical sense. A real fractal has infinite levels, which can only be revealed in the mathematical world, but a pre-fractal is a limited hierarchy indicating fractal-like geometric form, which can be found in any textbooks on fractals. We can use the ideas from fractal geometry to research pre-fractals, including regular pre-fractals and random pre-fractals. The coast of Britain can be regarded as a random pre-fractal curve instead of a real fractal line. However, we can study the coast of Britain using the ideas from fractals and fractal dimension. Similarly, cities are not true fractals, but proved to be random pre-fractals because urban form has no characteristic scales. A great number of empirical studies show that, based on certain scaling range, urban form satisfy three necessary and sufficient conditions for fractals (Table 1). Urban form follow power laws, which indicates that cities can be treated as pre-fractals. The basic property of a random pre-fractal object is that its scaling range is limited, and its fractal dimension value is based on the scaling range (see, e.g., Addison, 1997).

**Table 1 Three pre-conditions for understanding, developing, and generalizing fractal concepts**

| Conditions | Formula | Note |
|---|---|---|
| **Scaling law** | $\mathrm{T}f(x)=f(\lambda x)=\lambda^{b}f(x)$ | The relation between scale and the corresponding measures follow power laws |
| **Fractal dimension** | $d_{\mathrm{T}}<D<d_{\mathrm{E}}$ | The fractal dimension $D$ is greater than the topological dimension $d_{\mathrm{T}}$ and less than the Euclidean dimension of the embedding space $d_{\mathrm{E}}$. |
| **Entropy conservation** | $\sum_{i=1}^{N(r)}P_i^q r_i^{(1-q)D_q}=1$ | The Renyi entropy values of different fractal units (fractal subsets) are equal to one another. |

**Note**: T—scaling transform; $x$—scale variable; $f(x)$—a function of $x$; $\lambda$—scale factor; $b$—scaling exponent; $D$—fractal dimension; $d_{\mathrm{T}}$ --topological dimension; $d_{\mathrm{E}}$ --Euclidean dimension of embedding space; $q$—order of moment; $P_i$, $r_i$ —growth probability of the $i$th fractal set and its linear scale; $D_q$—generalized correlation dimension; $N(r)$—number of fractal units with linear size $r$; $i$—ordinal number: $i$=1, 2, 3, …, $N(r)$.

## 2.2 Fractal geometry: an approach to scale-free analysis

Fractal geometry is a powerful tool for scaling analysis of scale-free phenomena such as urban form. Scaling suggests that there is no characteristic scale in a thing. Cities, in many aspects, have no characteristic scale and cannot be effectively modeled by the conventional mathematical methods. In contrast, urban phenomena can be well characterized by fractal parameters. Natural and social phenomena can be roughly divided into two categories: one is the phenomena with characteristic scales, and the other is the phenomena without characteristic scales. The former can be termed *scaleful phenomena*, and the later can be termed *scale-free phenomena* (Table 2). For the scaleful phenomena, we can find definite length, area, volume, density, eigenvalue, mean value, standard deviation, and so on. If the spatial distribution of this kind of phenomena is converted into a probability distribution, it has clear and stable probability structure and thus can be described with Gaussian function, exponential function, logarithmic function, lognormal function, Weibull function, etc. The conventional higher mathematics can be used as an effective tool for modeling and analyzing such phenomena. On the contrary, for the scale-free phenomena, we cannot find effective length, area, volume, density, eigenvalue, mean value, standard deviation, and so forth. If the spatial distribution of this sort of phenomena is transformed into a probability distribution, it can be characterized with power functions, Cobb-Douglas function (production function), or some type of functions including hidden scaling. The probability structure of the scale-free distributions is not certain. Traditional advanced mathematics cannot effectively characterize such phenomena. In recent years, a number of theoretical tools for scale-free analysis are emerging, including fractal geometry, wavelet analysis, allometric theory, and complex network theory. Among various "new" tools, fractal geometry represents an excellent method for scale-free modelling and scaling analysis.

**Table 2 Two types of natural and social phenomena: scaleful and scale-free phenomena**

| Type | Probability distribution | Characteristics | Example | Mathematical tools | Description |
|---|---|---|---|---|---|
| **Scaleful phenomena (with characteristic scales)** | Normal, exponential, logarithmic, lognormal, Weibull, etc. | We can find definite length, area, volume, density, eigenvalue, mean value, standard deviation, and so on | Urban population density distribution, which follows exponential law | Traditional higher mathematics includes calculus, linear algebra, probability theory and statistics. | Entropy function and Gaussian distribution |
| **Scale-free phenomena (without characteristic scale)** | Power law, various hidden scaling distributions | We cannot find effective length, area, volume, density, eigenvalue, mean value, standard deviation, and so on | Urban traffic network density distribution, which follows power law | Fractal geometry, complex network theory, allometry theory, scaling theory | Fractal dimension and Pareto distribution |

A city is a complex system with multifaceted characteristics. In some aspects, a city has characteristic scales, e.g., urban population density distribution, which follows negative exponential law and can be described with Clark's model (Clark, 1951). The spatial distribution function can be derived from the principle of entropy maximization (Chen, 2008). In another aspect, a city has no characteristic scale, e.g., urban traffic network density distribution, which follows inverse power law and can be characterized with Smeed's model (Smeed, 1963). The corresponding spatial distribution can be characterized by spatial correlation and allometric scaling (Chen *et al*, 2019). Where land use is concerned, urban form follow power law distribution and can be treated as random pre-fractal patterns (e.g., Batty and Longley, 1994; Frankhauser, 1994). In this sense, we cannot find effective characteristic scales for urban morphology. Consequently, the traditional methods of quantitative analysis and mathematical modeling are often invalid for the research on urban form and growth. As a substitute, fractal geometry is one of feasible mathematical tools for the spatial analysis of cities.

### 2.3 How to define city fractals?

The angle of view for fractal studies of cities depends on the definition of embedding space. A city fractal based on digital maps or remote sensing images can be defined in a 2-dimensional embedding space, and also it can be defined in a 3-dimensional embedding space (Thomas *et al*, 2012). Generally speaking, fractal cities are defined in a 2-dimensional embedding space (e.g. Batty and Longley, 1994; Benguigui *et al*, 2000; Frankhauser, 1994). However, some scholars study fractal cities through 3-dimensional embedding space (e.g., Qin *et al*, 2015). The fractal city defined in three-dimensional embedding space has attracted the attention of geographers (Thomas *et al*, 2012). In fact, a fractal based on the 3-dimensional embedding space can be explored through the 2-dimensional embedding space. In the simplest case, the relationship between the fractal dimension based on 2-dimensional embedding space, $D^{(2)}$, and the fractal dimension based on 3-dimensional embedding space, $D^{(3)}$, is as follows, $D^{(3)}=1+ D^{(2)}$ (Vicsek, 1989).

For simplicity, we define the city fractals in a 2-dimensional embedding space. The main reasons are as follows.

**First, fractal dimension is used to replace the 2-dimensional urban area rather than the 3--dimensional urban volume.** In order to study a city, we must describe a city; in order to describe a city, we must know its basic measures such as population size, urban area, and economic output. Unfortunately, urban form has no characteristic scales due to its fractal properties, and thus urban boundary cannot be objectively determined. In this case, urban area cannot be objectively calculated because the measurement results depend on scales. This is the well-known scale-dependence property of urban form, the cause lies in scale-free distribution of urban land use. In this case, fractal dimension of urban form can be employed to replace urban area to reflect the extent of space filling. The fractal dimension as a degree of urban space filling is exactly a substitute of urban area. Urban area is a scale-dependent measure, while fractal dimension is scaleful parameter. In this sense, fractal dimension is more effective than urban area to reflect urban spatial development. By the way, some scholars prefer to define a city fractal in a 3-dimensional space, this means that they try to calculate a fractal dimension based on 3-dimensional embedding space to replace urban volume.

**Second, the general principle of model building is based on reduction of dimension.** The effective skill of scientific quantitative analysis is to reduce dimension instead of to increase dimension. The basic relation between spatial dimension $n$ and the degree of analytical complexity $C$ can be expressed as $C=n(n-1)/2$, which represents the least statistical parameter number for quantitative analysis. The well-known Clark's law of urban population density distribution in a 2-dimensional space is actually based on a 1-dimensional space modeling, but this model reflect the geographical information in a 3-dimensional space (Clark, 1951). In other word, the population distribution in the 3-dimensional space is projected to the 2-dimensional space by population density, and then the mathematical expression is established on the basis of the 1-dimensional space with the help of statistical averaging (Chen, 2008). The same is the case with Smeed's model on urban traffic density distribution (Batty and Longley, 1994; Chen *et al*, 2019; Smeed, 1963). If we study a city fractal through a 3-dimensional embedding space, the amount of work and difficulty of fractal dimension calculation is considerably increased, and the accuracy of fractal parameter estimation is reduced, but the increment of gained geographic information is very limited. In short, it is hard to promote the analytical effect of fractal cities significantly by substituting the 2-dimensional embedding space with the 3-dimensional embedding space.

**Third, the allometric scaling relation between population and land use suggests that urban form should be defined in a 2-dimensional space.** The allometric scaling exponent $b$ is the ratio of the fractal dimension of urban form $D_f$ to the dimension of urban population $D_p$, that is, $b=D_f/D_p$. Empirical studies show that the $b$ values are close to 0.85 (Chen, 2010). If $D_f>2$, then we have $D_p>2/0.85=2.35$. Based on Clark's law and scaling analysis, urban population distribution proved to be a 2-dimension phenomena ($D_p=2$) (Chen and Feng, 2012). If the urban form is defined in a 3-dimensional embedding space, the fractal dimension $D_f$ values will come between 2 and 3, and the allometric scaling exponent $b$ values will be greater than 1. However, the observational values of allometric scaling exponent $b$ values range from 2/3 to 1 in the most cases, that is, $2/3<b<1$ (Chen, 2010; Lee, 1989; Louf and Barthelemy, 2014a). This suggests that the dimension of urban form, $D_f$, comes between 1 and 2. In fact, in urban studies, fractal dimension is not a concept of comparability. The fractal dimension value depends on the definition of embedding space.

If a city fractal is defined in a 2-dimensional embedding space, the fractal form includes two aspects: urban area and urban boundary. The above discussion is actually based on urban area, but urban boundary can be treated fractal lines (Batty and Longley, 1988; Batty and Longley, 1994; Benguigui *et al*, 2006; Chen, 2011; Longley and Batty, 1989a; Longley and Batty, 1989b; De Keersmaecker *et al*, 2003). The closed urban boundary curve is termed *urban envelope*, in which we can determine a Euclidean urban area (Batty and Longley, 1994; Longley *et al*, 1991). The length of urban boundary and the Euclidean area within the urban envelope follow the geometric measure relation as follows

$$A = aL^{2/D_b}, \tag{1}$$

where $A$ refers to the Euclidean area of a city (urban area), $L$ denotes the length of urban envelope (urban perimeter), $a$ is the proportionality coefficient, and $D_b$ is the fractal dimension of urban boundary, which can be termed boundary dimension (Chen, 2011). In fact, equation (1) can be generalized to the more general expression as below (Benguigui *et al*, 2006; Chen, 2013):

$$A = aL^{D_f/D_b}, \tag{2}$$

where $A$ refers to the Euclidean area of a city, and $D_f$ is the fractal dimension of "urban area". Equation (2) is in fact an allometric scaling relation of urban shape (Chen, 2013). The topological dimension of urban boundary is $d_T^b=1$, so the boundary dimension is greater than 1. The fractal parameter value comes between 1 and 2, that is, $1< D_b <2$. Now, a question appears. What determines the lower limit of fractal dimension of urban morphology, urban area or urban boundary? The answer is clear. If we study urban form and try to substitute urban area with form dimension, it is the topological dimension of urban area that determines the least value of the fractal dimension; on the other, if we research urban boundary and attempt to replace urban perimeter length with boundary dimension, it is the topological dimension of urban boundary that determine the minimum value of the fractal dimension. In the most cases, we study urban impervious area which is represented by the pixels of buildings (fractal separated spaces) rather than urban boundary (fractal lines).

## 2.4 The lower and upper limits of fractal dimension

Fractal dimension values have strict lower limit and upper limit. However, what are the lower limit and upper limit of the fractal dimension of urban from? This is still a pending question. Empirically, if a city fractal is defined in a 2-dimensional embedding space, the fractal dimension value come between 0 and 2 (Chen, 2012; Chen, 2018; Shen, 2002; Thomas *et al*, 2007). In theory, the lower and upper limits of fractal dimension of urban form rely on the topological dimension and embedding dimension. In many cases, the box-counting method is employed to estimate the fractal dimension values of urban form. The lower limit of the fractal dimension $D_{min}$ depends on the topological dimension of urban form $d_T$, while the upper limit $D_{max}$ depends on the Euclidean dimension of the embedding space $d_E$. As indicated above, the embedding space can be defined as a 2-dimensional space, thus the Euclidean dimension of $d_E=2$, so we have $D_{max}\leq d_E=2$. As for the topological dimension of urban form, $d_T$, in theory, it should be $d_T=0$. Therefore, we have $D_{min}\geq d_T=0$.

How to determine the topological dimension of urban form? As we know, the Lebesgue measures of real fractals are zero (Mandelbrot, 1982). This suggests that, if we treated urban form as a fractal, the urban area of land use should be treated as zero. Please note that this is based on theoretical understanding, which is different from reality. How to understand the assumption that the area of a city fractal is 0? This means that an urban fractal can be reduced to either a separated space or a space-filling curve under the limit conditions. For a separated space, the topological dimension is $d_T=0$; while for a space-filling curve, the topological dimension is $d_T=1$. In fact, using ArcGIS technique, we can reduce a city fractal to a separated space rather than a space-filling curve. A separated space of a city comprises pixels or building cells on a remote sensing image or digital map. This indicates that the topological dimension of city fractals is $d_T=0$ instead of $d_T=1$. According to Shen (2002), the box dimension values of Baltimore come between 0.6641 and 1.7211 from 1792 to 1992 year.

In practice, the lower and upper limits of fractal dimension of urban form depend on the methods of defining study area. There are two approaches to obtaining the time series of the fractal dimension values of urban growth and form (Chen, 2012). One is based on constant study area (Batty and Longley, 1994; Shen, 2002), and the other, based on variable study area (Benguigui *et al*, 2000; Feng and Chen, 2010). Each approach has its advantages and disadvantages (Table 3). If we define a study area with fixed size for different years, the largest box can be determined by the urban boundary of the most recent year. Then, the largest box can be applied to digital maps of the city in previous years (Figure 1(a)). Using the same set of boxes, we estimate the fractal dimension values of urban form in different years. Based on this approach, the fractal dimension values of a city's form in different years are more comparable. The time series of fractal dimension values can better reflect space replacement process of urban region. The subsets of the time series are termed sample paths. If a sample path is very long, the original urban form can be treated as a point. As a result, the fractal values may come between 0 and 2 (Chen, 2012; Thomas *et al*, 2007). In contrast, if we define a variable study area, the size of the largest box is determined by the urban boundary in a given year. Thus, the largest boxes are different from year to year (Figure 1(b)). Based on this approach, the comparability of fractal dimension values of urban form in different years is reduced. But these fractal dimension values can better reflect the degree of urban space filling. As a result, the fractal values may come between 1 and 2 (Chen, 2012).

**Table 3 Two approaches to defining the study area for fractal dimension estimation of urban form**

| Approach | Property | Merit | Demerit | Dimension range |
|---|---|---|---|---|
| **Constant study area** | Fixed size | The comparability of fractal parameters of different years is strong. The time series of fractal dimension can be used to reflect space replacement of urban region. | The reality of fractal parameters of each year is weak. | Come between 0 and 2 |
| **Variable study area** | Unfixed size | The reality of fractal dimension values of urban form is strong. The time series of fractal dimension can be used to reflect space filling of urban growth. | The comparability of fractal parameters of different years is weak. | Come between 1 and 2 |

# 3. Fractal modeling of urban form

## 3.1 Two research directions of fractal cities

A complete scientific research process comprises two elements. One is to describe a system, and the other is to understand the mechanism of the system's work. In short, scientific studies should proceed first by describing how things work and later by understanding why (Gordon, 2005). Accordingly, scientific method contains two elements: description and understanding. Concretely speaking, as stated by Henry (2002, p14): "The two main elements of this scientific method are the use of mathematics and measurement to give precise determinations of how the world and its parts work, and the use of observation, experience, and where necessary, artificially constructed experiments, to gain understanding of nature." A comparison between the two elements of scientific process can be drawn as follows (Table 4). The most important method of scientific description is to establish mathematical models.

**Table 4 A complete scientific research process consists of two elements**

| **Element** | **Level** | **Method** | **Purpose** | **Result** | **Finding** | **Fractal theory** |
|---|---|---|---|---|---|---|
| **Description** | Macro level | Mathematics, measurement, and computation | Data, numbers | Show characters of a system's behavior | How a system works | Geometrical method |
| **Understanding** | Micro level | Observation, experience, experiments, and simulation | Insight, sharpen questions | Reveal dynamical mechanism | Why the system works in this way | Ideas of complex systems |

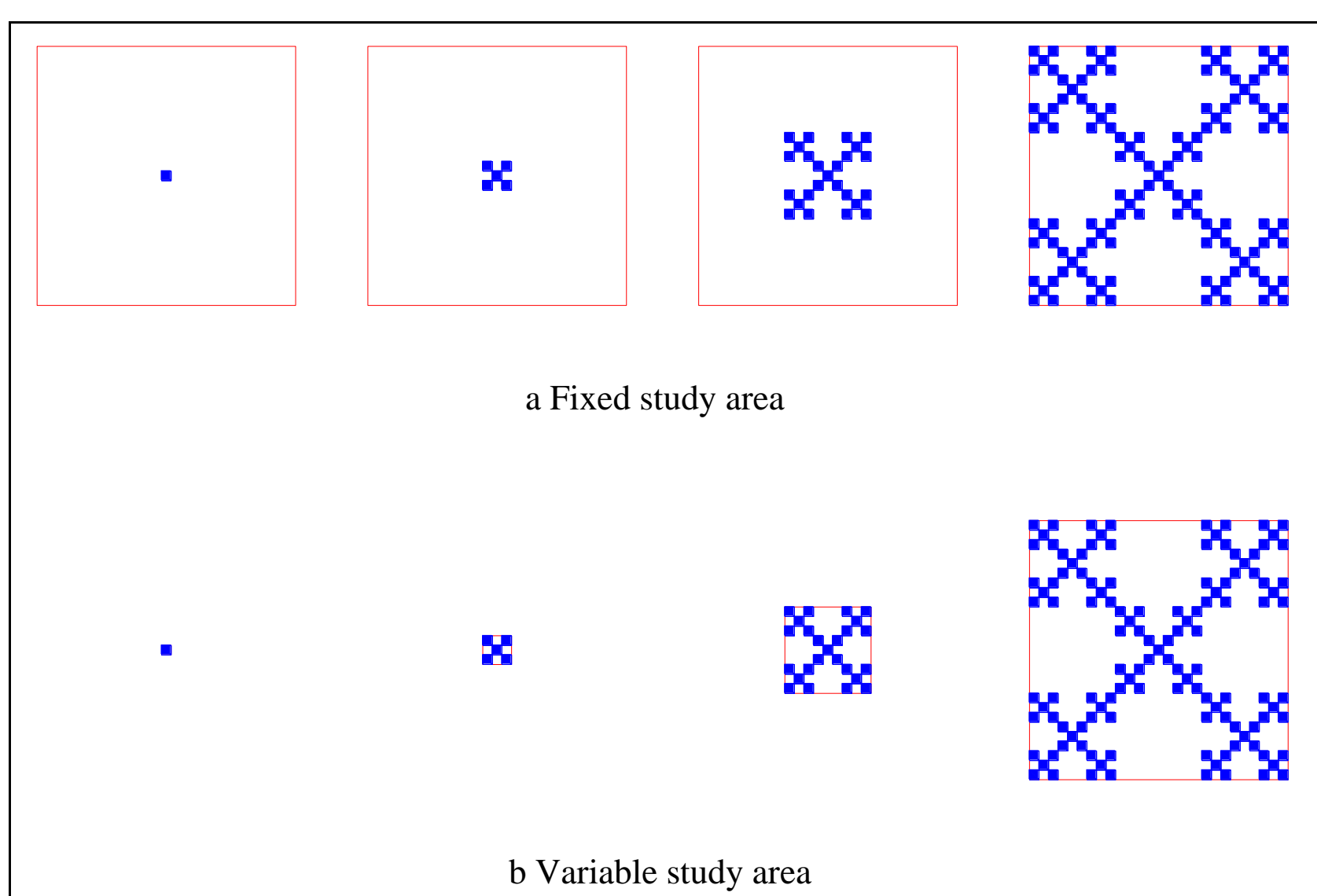

**Figure 1 The sketch maps for two types of approaches to defining study areas for fractal dimension estimation of urban form (by Chen, 2012)**

**Note:** The square frames surrounding the growing fractals represent the study area of fractal dimension measurements. Figure 1(a) shows a fixed study area, and Figure 1(b) displays a variable study area, the size of which depends on the extent of fractal city cluster.

Fractal theory comprise two related parts: one is the scaling theory of complex systems, and the other is the mathematical method known as fractal geometry. As a complex system theory, it can be employed to understand complexity of cities; as a geometry, it can be used to describe cities from the angle of view of scaling analysis. In fact, a mathematical theory plays two roles in any scientific research (Table 5). One is make models and develop theory (mathematical modeling), and the other is process experimental and observational data (statistical analysis). In urban studies, fractal geometry can serve two functions. One is to establish models for cities and systems of cities, and the other is to make empirical analysis of cities using observational data. Many scholars utilize fractal geometry to process the observational data of urban geography, but I emphasize the basic function: mathematical modeling. No matter what types of studies are made, there is no contradiction between models and observed data. All models relies heavily on observational data.

**Table 5 Two functions of fractal geometry in urban studies**

| Function | Use | Purpose | Approach |
|---|---|---|---|
| **Theoretical** | Present postulates and make models | Develop urban theory | Build mathematical models based on fractals or fractal dimension |
| **Empirical** | Process experiment and observational data | Solve practical problems in reality | Rely heavily on fractal dimension |

In fact, one of the main tasks in scientific research is to make models. As Neumann (1961) said: "The sciences do not try to explain, they hardly even try to interpret, they mainly make models." I agree with Hamming (1962), who said: "The purpose of modelling is insight, not numbers." Karlin (1983) has similar viewpoint: "The purpose of models is not to fit the data, but to sharpen the questions." However, the confidence level of a model depends heavily on the relationship between mathematical expression and observed data. In order to verify a mathematical model, we must fit it to observational data and illustrate the statistical relationships and analytical effect. I am very much in favor of his viewpoint of Louf and Barthelemy (2014b), who said: "The success of natural sciences lies in their great emphasis on the role of quantifiable data and their interplay with models. Data and models are both necessary for the progress of our understanding: data generate stylized facts and put constraints on models. Models on the other hand are essential to comprehend the processes at play and how the system works. If either is missing, our understanding and explanation of a phenomenon are questionable. This issue is very general, and affects all scientific domains, including the study of cities." The basic functions of mathematical models are explanation and prediction. As Fotheringham and O'Kelly (1989) pointed out, "All mathematical modelling can have two major, sometimes contradictory, aims: explanation and prediction." Not only that, as Kac (1969) observed, "The main role of models is not so much to explain or predict—although ultimately these are the main functions of science—as to polarize thinking and to pose sharp questions." The chief uses of fractal models lie in explanation and prediction. Let's take the logistic model of fractal dimension growth curves as an example. The model can be used to explain the speed change characteristics of urban growth (Chen, 2018). It can tell us when the growth rate of a city will peak. It can also tell us the maximum space-filling index of a city's land use. What is more, the model can sharpen questions for us. For example, the similarity and difference between the model of fractal dimension growth curves of Chinese cities such as Beijing and that of the cities in western countries such as London, Baltimore and Tel Aviv give rise to new thinking about the spatial dynamics of urban evolution.

### 3.2 Two approaches to modeling cities

As indicated above, one of important tasks of fractal urban studies it to make models. As Longley (1999, Page 605) pointed out, "In the most general terms, a 'model' can be defined as a 'simplification of reality', nothing more, nothing less." In scientific research, mathematical models can be classified into two categories: *mechanistic models* and *parametric models* (Su, 1988). Accordingly, there exist two approaches to establishing mathematical models: *analytical method* and *experimental method* (Zhao and Zhan, 1991) (Table 6). The so-called analytical method is the approach to deriving a mathematical model with the help of the existing scientific theories and laws, and in light of the relationship and evolution of the various components of the studied system. The process is as follows: establish a functional equation based on one or more postulates, and then find the general solution to the functional equation. The solution to the equation is exactly the theoretical model (mechanistic or structural model) we need. The experimental method is to select the most approximate model in a set of hypothetical or imaginary models so that the model can be well fitted to the observational or experimental data. What is more, the model will not give rise to logical contradiction and difficulty in interpretation. Thus we have an empirical model (parametric or functional model). In geography, the traditional gravity model is an empirical model, which is obtained by analogy with Newton's law of universal gravitation. In contrast, the spatial interaction model of Wilson (1968) is a theoretical model. The model is derived by constructing the postulates and solving the maximum entropy equation of traffic flows. The two types of models are not opposed, but can be transformed into each other. An effective theoretical model must be an empirical model, which must be well fitted to observation data. On the other hand, an empirical model will become a theoretical model by mathematical demonstration. A typical example is Clark's urban population density model (Clark, 1991). The model was originally presented as an empirical model based on observation data (Batty and Longley, 1994). However, it has become a theoretical model because it can be derived from the postulates of spatial entropy maximization of urban population distribution (Chen, 2008). In an article, limited to the conditions at the time, we may fulfil some aspect of the research work, not necessarily complete all the research process.

**Table 6 Two types of models and methods of model building**

| Model type | Property | Building method | Principle | Example |
|---|---|---|---|---|
| Mechanistic model (structural model) | Theoretical model | Analytical method | Postulates and demonstration | Wilson’s spatial interaction model |
| Parametric model (functional model) | Empirical model | Experimental method | Data and fitting | Traditional gravity model |

## 3.3 Fractal models and parameters of cities

We have at least three approach to develop mathematical models of urban form by using ideas from fractal theory. The first is to make new models, the second is to improve the old models, and the third is to borrow models from other disciplines (Table 7). A typical example is the models of fractal dimension growth curve of urban form, different approaches result in different models, and different models are suitable for different situations (Chen, 2012; Chen, 2018). It is necessary to briefly comment on the third way. In scientific research, a mathematical model can be transplanted from a field and applied to another field. The logistic function was originally proposed by Verhulst in 1838 to prediction population growth (Banks, 1994). Today, the well-known logistic function has been employed to predict many growing phenomena in many different fields, including urbanization level and fractal dimension growth (Chen, 2018). Similarly, Boltzmann equation can also be generalized to other fields and to model urban growth (Benguigui *et al*, 2001; Chen, 2012). The allometric growth equation of urban geography came from biology (Naroll and Bertalanffy, 1956; Chen, 2011). The gravity model of geography resulted from Newton's law of universal gravitation by analogy, and the spatial autocorrelation models of geography come from mathematical and statistical biology. These examples are too numerous to enumerate. The uniqueness of different fields is always determined by the physical meaning of model parameters rather than by the expression of mathematical models. The mathematical expression of a model is often general, but the parameters are for special purposes. The same mathematical model can be applied to many different fields, but different fields have different parameter meanings.

**Table 7 Three approaches to develop models for fractal dimension growth curves of urban form**

| Approach | Example and mathematical expression | Name |
|---|---|---|
| Make new models | $D(t)=\dfrac{D_{\max}}{1+(D_{\max}/D_{(1)}-1)t^{-b}}$ <br> $D(t)=D_{\min}+\dfrac{D_{\max}-D_{\min}}{1+[(D_{\max}-D_{(1)})/(D_{(1)}-D_{\min})]t^{-b}}$ | Growth function of hidden scaling |
| Improved old model | $D(t)=\dfrac{D_{\max}}{1+(D_{\max}/D_{(0)}-1)e^{-kt^{2}}}$ | Quadratic logistic function |
| Borrow model from other discipline | $D(t)=D_{\min}+\dfrac{D_{\max}-D_{\min}}{1+[(D_{\max}-D_{(0)})/(D_{(0)}-D_{\min})]e^{-kt}}$ | Boltzmann equation |

**Note**: (1) Models. The logistic function and Boltzmann equation of fractal dimension growth curve were demonstrated by Chen (2012), and the quadratic logistic function was derived and demonstrated by Chen (2018) and Chen and Huang (2019). (2) Parameters. $D(t)$—fractal dimension of urban form at time $t$; $D_{(0)}$—the initial value of fractal dimension of urban form ($t$=0); $D_{\max}$, $D_{\min}$—the upper limit and lower limit of fractal dimension; $b$—the scaling exponent of fractal dimension growth; $r$—the original growth rate of fractal dimension.

The notion of maximum and minimum of fractal dimension discussed above is important for making models of the fractal dimension growth curves of urban form. The fractal dimension growth curve results from the time series of urban growth. In theory, we can calculate the fractal dimension values of a city's form in different times. This values compose a sample path of fractal dimension, and further form a curve of fractal dimension change of urban morphology. A sample path can be regarded as a subset of a time series (Diebold, 2007). Due to the lower and upper limits of urban fractal dimension, a fractal dimension growth curve takes on squashing effect and can be described with one of sigmoid functions such as logistic function and Boltzmann's equation (Chen, 2012; Chen, 2014; Chen, 2018). On the other hand, how to determine fractal parameter values depends on specific research objectives and data conditions. This is a complex problem and needs to be judged on the basis of long-term research experience. Even for theoretical research, if the sample path of fractal dimension is short, we can take $D_{min}$=1 and adopt the quadratic Boltzmann equation. For example, in one of the studies made by Chen (2018), the time span is about 25 years (1984-2008). All the fractal dimension values are greater than 1. On the other hand, even for application research, if the sample path of fractal dimension is very long, we can take $D_{min}$=0 and adopt the quadratic logistic function. For instance, for the study of Shen (2002), the time span is about 200 years (1792-1992). One of the fractal dimension values for early years is less than 1. The situations can be classified into four groups and tabulated as below (Table 8).

**Table 8 Four cases for the lower limit of fractal dimension growth curves of urban form**

| | **Fixed study area** | **Variable study area** |
|---|---|---|
| **In theory** | $D_{min}$=0, logistic function | $D_{min}$=0, long sample path, logistic function;<br>$D_{min}$=1, usual cases, Boltzmann equation |
| **In practice** | $D_{min}$=1, short sample path, Boltzmann equation;<br>$D_{min}$=0, usual cases, logistic function | $D_{min}$=1, Boltzmann equation |

# 4. Questions and discussion

## 4.1 Problems of fractal dimension values

The concept of fractal dimension proceeded from Hausdorff's fractional dimension. Today, there various definitions for fractal dimension, and the common fractal dimensions in urban studies is box dimension and similarity dimension. The box dimension is mainly suitable for the spatial structure of urban form and systems of cities, while the similarity dimension is chiefly applied to urban hierarchies, including hierarchies of cities and hierarchies of urban internal elements such as land use patches. Generally speaking, fractal dimension values come between the topological dimension and the Euclidean dimension of embedding space. For a regular fractal, if fractal copies/units have no overlapping, its Hausdorff dimension will equal similarity dimension. Empirically, both Hausdorff dimension and similarity dimension can be represented with box dimension. All these dimension values are less than the Euclidean dimension of the embedding space and greater than the topological dimension of fractal objects. However, if fractal copies have overlapped parts, the similarity dimension will exceed the dimension of embedding space in value. Thus, similarity dimension will not equal Hausdorff dimension or box dimension. In contrast, the box dimension will never exceed the embedding dimension.

Let's examine two kinds of fractal dimension of the fractals with overlapped parts. The interior boundary line of the Sierpinski gasket is a typical fractal line with overlapped parts (Figure 2). The initiator is a straight line segment with length of unit (Figure 3(a)), the generator is a curve consisting of 5 straight line segments with length of 1/2 unit (Figure 3(b)). From step 3 on, fractal copies begin to overlap one another, and the overlapped parts are marked with red circles (Figure 3(c), Figure 3(d)).

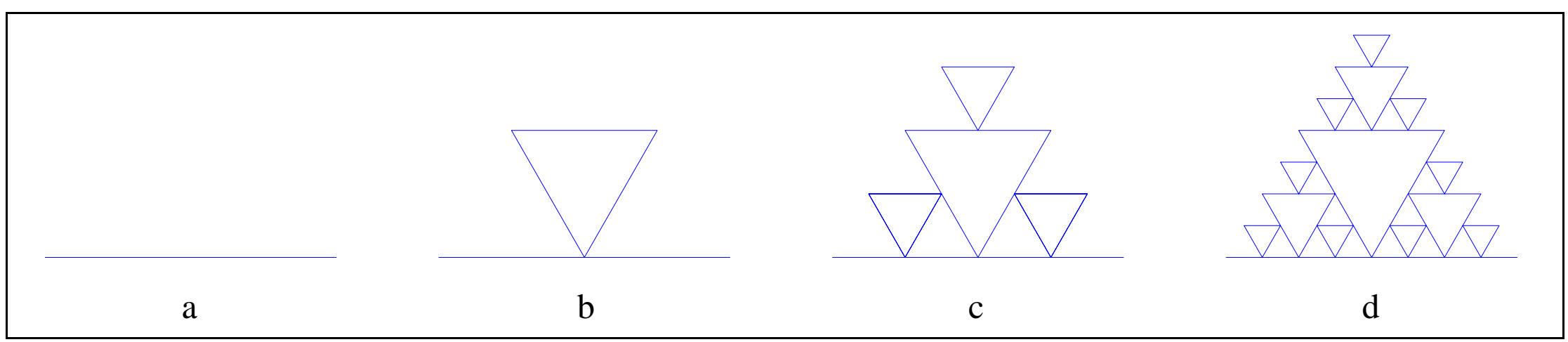

**Figure 2 The interior boundary line of the Sierpinski gasket (The first four steps)**

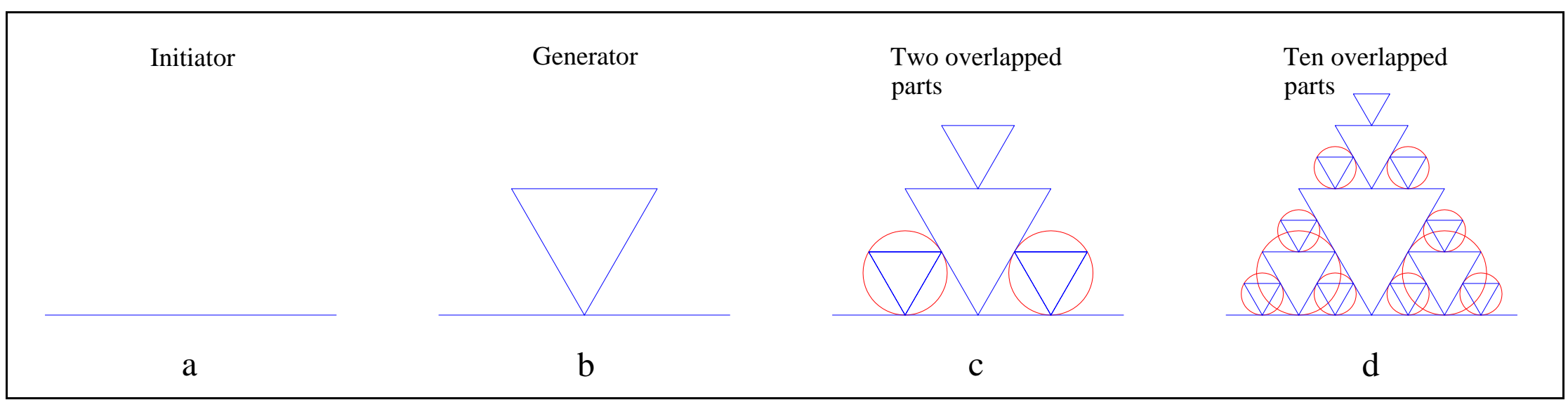

**Figure 3 A special fractal line with overlapped parts (The first four steps)**

The similarity dimension and box dimension can be calculated by the ideas from fractal dimension. In the *m*th step, the length (linear size) of line segments can be expressed as

$$s_m = (\frac{1}{2})^{m-1}, \tag{3}$$

where $m$=1,2,3,…denotes the ordinal numeration of steps. The number of line segments in each step can be counted by two different ways. One is to repeat counting the overlapped parts, and the other is to count the overlapped parts only one times. For example, for the curve of step 3 (Figure 2(c), Figure 3(c)), the number of line segments is $N_3=5^2=25$ according to the first counting way, and $N_3$=3*5+2$^2$=19 according to the second counting way. According to the first way with repeated counting, the line segment number in the $m$th step is

$$N_m = 5^{m-1}. \tag{4}$$

Thus the *similarity dimension* is

$$D_{\mathrm{s}} = -\frac{\ln(N_{m+1}/N_m)}{\ln(s_{m+1}/s_m)} = \frac{\ln 5}{\ln 2} = 2.322 > d = 2. \tag{5}$$

According to the second way without repeated counting, the line segment number of step $m$ is

$$N_m = 3N_{m-1} + 2^{m-1}, \tag{6}$$

where $N_0$=0 for $m$=1. By recurrence, we have

$$N_m = \sum_{j=0}^{m-1}(3^{m-1-j}2^j) = 3^{m-1}\sum_{j=0}^{m-1}[(\frac{2}{3})^j], \tag{7}$$

where $j$=1,2,…$m$-1. Under the condition of limit, the result is

$$N_m = \lim_{m\to\infty}\left[3^{m-1}\sum_{j=0}^{m-1}[(\frac{2}{3})^j]\right] = 3^{m-1}\frac{1}{1-2/3} = 3^m. \tag{8}$$

This suggests that when $m$ becomes large enough, $N_m$ will approaches $3^m$. So the *box dimension* is

$$D_{\mathrm{b}} = -\frac{\ln N_m}{\ln s_m} = \frac{m\ln 3}{(m-1)\ln 2} \xrightarrow{m\to\infty} \frac{\ln 3}{\ln 2} \approx 1.585 < d = 2. \tag{9}$$

For this special regular fractal, box dimension equals Hausdorff dimension in theory. Therefore, for the regular monofractals with overlapped units, we have the following relation: topological dimension < Hausdorff dimension = box dimension <embedding space dimension <similarity dimension. However, for the regular monofractals without overlapped units, the dimension relation is as follows: topological dimension < Hausdorff dimension = box dimension= similarity dimension <embedding space dimension.

The phenomenon of overlapped fractal units resulting in fractal dimension values greater than embedding space dimension can be employed to explain abnormal multifractal spectral curves. In theory, the generalized correlation dimension, $D_q$, should come between 0 and 2 if a fractal city is defined in a 2-dimensional embedding space. However, in many cases, the generalized correlation dimension values of urban morphology always exceed 2 or even go beyond 3 if the moment order, $q$, approaches negative infinity (Chen and Huang, 2018). The reason is that, even based on the box-counting method, if $q < 0$ or $q > 1$, we will get similarity dimension instead of strict box dimension of urban form. When $q < 0$, it means that the small patches in urban pattern are zoomed in gradually, and this leads to the overlapping and interlacing of random fractal units. If and only if the urban spatial structure is very orderly organized, the overlapping distributions of the magnified patches will be reduced to be omitted. In this sense, multifractal spectrums can be adopted to appraise the quality of spatial structure of cities and systems of cities.

## 4.2 Spatial meanings of fractal dimension

Fractal dimension represents a characteristic parameter for scale-free phenomena, which have no characteristic lengths and cannot be effectively described by traditional mathematical methods. Where cities is concerned, the meanings and uses of fractal dimension of urban form rest with at least three aspects: *degree of space filling*, *degree of spatial uniformity*, *degree of spatial complexity*. As a space-filling index, fractal dimension can be used to reflect the replacement process of urban and rural space in theory. Unfortunately, it is both impossible and unnecessary to distinguish between urban area and rural area strictly. When we define a study area for a fractal cities, it comprises urban buildings, rural buildings, and other types of land. Various types of land form a hierarchy with cascade structure of land use based on different levels of scales (Kaye, 1989). In the urban regions, there are rural buildings, and in the rural regions, there are urban buildings. If we examine a city's form from various spatial scales, we can find interlaced distributions of urban and rural land and buildings. The hierarchy with cascade structure of urban and rural landscapes should be described with multifractals (Chen, 2016). To solve the problem, we can use the concepts space-filling extent, $U(t)$, and space-saving extent, $V(t)$, to replace urban land use and rural land use (Chen, 2012).

In generalized correlation dimension spectrum, three parameters are very important, that is, capacity dimension, information dimension, and correlation dimension. Among the three common parameters, capacity dimension is the basic one. The essence of capacity dimension is just space-filling ratio, and this can be demonstrated easily. Space-filling measures should be defined by logarithmic scales rather than conventional scales. The reason is that spatial recursion process is based on exponential decay and logarithmic scale (Batty and Longley, 1994; Chen, 2016; Goodchild and Mark, 1987). Let's define an index of space filling as follows

$$F=\frac{\ln A_{\mathrm{b}}(r)}{\ln A(r)}=\frac{\ln N_{\mathrm{b}}(r)}{\ln N(r)}, \quad (10)$$

where $F$ denotes the space-filling ratio, $A_b$ refers to filled area, which can be represented by impervious area, $A$ is the total area, $N_b$ is the number of nonempty boxes, and $A$ is number of all boxes. It can be proved that

$$2F=\frac{2\ln A_{\mathrm{b}}(r)}{\ln A(r)}=\frac{2\ln N_{\mathrm{b}}(r)}{\ln N(r)}=D_0, \quad (11)$$

where $r$ denotes the ordinal numeration of steps, and $D_0$ refers to capacity dimension. For example, the growing fractal displayed in Figure 1, we have

$$2F=\frac{2\ln(5^m)}{\ln(9^m)}=\frac{2\ln(5)}{\ln(9)}=\frac{\ln(5)}{\ln(3)}=D_0, \quad (12)$$

where $m$ refers to the ordinal numeration of steps. This suggests that the doubling space-filling ratio yields capacity dimension of a regular fractal. This conclusion can be generalized o urban morphology. On the other hand, in equation (10), the numerator is Hartley entropy, $H$, and the denominator is the maximum entropy, $H_{\max}$ (Chen and Huang, 2018). If the minimum entropy and minimum fractal dimension are zero, the space-filling ratio proved to be the normalized entropy and to the normalized capacity dimension (Chen *et al*, 2017). What is more, fractal dimension is proved to the scaling exponent of spatial correlation, and a correlation function can be expressed as

$$C(r)=C_1 r^{2(D_0-d)+1}, \quad (13)$$

where $r$ refers to distance, $C(r)$ denotes spatial correlation function, $C_1$ is proportionality coefficient, and $d$ represent embedding space dimension (Chen, 2013). Spatial correlation suggests spatial displacement, which corresponds time lag and implies spatial complexity. In short, fractal dimension means space filling, spatial uniformity, and spatial complexity (Table 9).

**Table 9 The three basic meanings of fractal dimension of urban morphology**

| Basic measurement | Principle | Meaning | Explanation |
|---|---|---|---|

| Degree of space filling | $2F=\frac{2\ln A_b(r)}{\ln A(r)}=\frac{2\ln N_b(r)}{\ln N(r)}=D_0$ | Capacity dimension equals doubled space-filling ratio | The space filling ratio equals the logarithm of occupied area divided by the logarithm of total area |
|---|---|---|---|
| Degree of spatial uniformity | $2F=\frac{2\ln H}{\ln H_{max}}=\frac{2\ln N_b(r)}{\ln N(r)}=D_0$ | Capacity dimension equals doubled normalized Hartley entropy | Entropy is a measure of spatial uniformity |
| Degree of spatial complexity | $C(r)=C_1 r^{2(D_0-d)+1}$ | Capacity dimension suggests a spatial correlation exponent | Spatial correlation indicates spatial complexity of cities |

**Note**: The formula of space filling degree is derived in this paper, and the spatial correlation function was presented by Chen (2013). About the relationships between entropy and fractal dimension, see Chen *et al* (2017).

### 4.3 Statistical evaluation of fractal parameters

It is necessary to discuss fractal dimension measurement methods and related statistical test parameters simply. In practice, the double logarithmic linear regression based on least square method can be employed to estimate fractal dimension values. Two methods can be utilized to make regression analysis: one is fixed intercept to 0, and the other, let intercept free. The former can be termed fixed intercept regression, and the latter can be termed free intercept regression. For theoretical analysis, the intercept should be fixed to 0 so that the proportionality coefficient of the corresponding fractal model equals 1. For positive studies, the intercept depends on the measurement results and should not be fixed to certain value (Huang and Chen, 2018). No matter which method is adopted, statistical test should be carried out for the calculation results (Table 10). The basic and most important statistic for fractal dimension test is goodness of fit, $R^2$, which is also termed determination coefficient. Actually, the $R$ statistic is called multiple correlation coefficient, which equals the absolute value of Pearson correlation coefficient for univariate linear regression analysis. Sometimes, we examine standard error and probability value, i.e., $P$ value, of a fractal dimension. A proper statement in scientific research should be presented by *confidence statement* (Chen *et al*, 2017). A confidence statement comprises two elements: *margin of error* and *level of confidence* (Moore, 2009). According to the standard error, $\delta$, we can estimate the margin of error of a fractal dimension; and according to the $P$ value, we can calculate the level of confidence of the fractal dimension. In most cases, the fractal dimension calculation is based on univariate linear regression analysis. For univariate linear regression, the $R^2$ value, the *F* statistic, *t* statistic, and the corresponding *P* value are equivalent to one another. What is more, the fractal dimension *D* and the $R^2$ value can be associated with the standard error *δ*. The formulae are as follows (see Appendix)

$$F = t^2 = \frac{vR^2}{1-R^2}, \tag{14}$$

$$\delta = D\sqrt{\frac{1/R^2-1}{v}}, \tag{15}$$

where *v* denotes degree of freedom. If the intercept of the log-log linear model for regression analysis is free, the degree of freedom is $v=n-2$; If the intercept is fixed to 0, the degree of freedom is $v=n-1$. Here *n* is the sample size, i.e., the data point number. Then, using the *t* distribution function tdist, we can convert the *t* statistic into the corresponding *P* value by means of MS Excel. The grammar is "=tdist(abs(*t* value), *v*, 2)". Thus based on the 95% level of confidence, the margin of error of the fractal dimension value can be approximately expressed as $D\pm2s$. This means that, in the absence of special requirements, the $R^2$ value will provide enough numerical information for statistical description of fractal dimension.

**Table 10 The transformation relationships between *F* statistic, *t* statistic, *P* values, standard deviation, and fractal dimension**

| Item | Free intercept (Arbitrary value) | Fixed intercept (Zero) |
|---|---|---|
| Fractal model | $N(r)=Kr^{-D}$ (0<*K*<2) | $N(r)=Kr^{-D}$ (*K*=1) |
| Logarithmic linear relation | $\ln N(r)=\ln K-D\ln r$ | $\ln N(r)=-D\ln r$ (ln*K*=0) |
| Degree of freedom, *v* | $v=n-2$ | $v=n-1$ |
| *F* statistic, *F*, *t* statistic, *t*, and goodness of fit, $R^2$ | $F=t^2=\frac{(n-2)R^2}{1-R^2}$ | $F=t^2=\frac{(n-1)R^2}{1-R^2}$ |
| Standard error *δ*, fractal dimension *D*, and $R^2$ | $\delta=D\sqrt{\frac{1/R^2-1}{n-2}}$ | $\delta=D\sqrt{\frac{1/R^2-1}{n-1}}$ |
| Margin of error of fractal dimension *D* (significance level *α*=0.05) | $D\pm2\delta$ | $D\pm2\delta$ |
| Excel conversion formula from *t* statistic to *P* value | $\text{tdist}(\text{abs}(t),n-2,2)$ | $\text{tdist}(\text{abs}(t),n-1,2)$ |
| Definition of *R* statistic | Pearson correlation coefficient | Cosine coefficient |

**Note**: (1) Fomulae. See Appendix. (2) Parameters. *r*—spatial measurement scale such as linear size of box; *N*(*r*)—

spatial measurement with linear size *r* such as the number of nonempty boxes; *K*—proportionality coefficient; *D*—fractal dimension; ln –natural logarithm function; *n*—sample size (data point number); *F*—*F* statistic; *t*—*t* statistic; *R*—multiple correlation coefficient; tdist, abs—MS Excel functions for *t* distribution and absolute value.

The analytical process and discussion of this paper is based on the standard definition of fractals. A fractal has three elements, i.e., *form*, *chance*, and *dimension* (Mandelbrot, 1977). The first definition of Mandelbrot (1982, page 15) based on dimension and chance is as follows: "A fractal is by definition a set for which the Hausdorff -Besicovitch dimension strictly exceeds the topological dimension." The second definition based on form and chance is as below: "A fractal is a shape made of parts similar to the whole in some way." The second definition is given by Mandelbrot but published by Feder (1988, page 11). The quantitative criterion of fractals is Hausdorff -Besicovitch dimension. Recent years, Jiang and his co-workers tried to relax the definition of fractals and give the third definition as follows: *A set or pattern is fractal if the scaling of far more small things than large ones recurs multiple times* (Jiang and Yin, 2014). According to the new definition, the quantitative criterion of fractals is replaced by the head/tail index (Jiang, 2013; Jiang, 2015): *the ht-index of a fractal set or fractal pattern is at least three* (Jiang and Yin, 2014). The new definition and criterion of fractals are very interesting and instructive. Sometimes, definitions of concepts or terms are most likely to lead to ambiguity, misunderstanding, and controversy. Therefore, scientific studies should sidesteps the terminological minefield so that we can move beyond the semantic debate (Gallagher and Appenzeller, 1999). On the one hand, we should leave certain room for developing and consolidating a definition as the research approach continues to mature (Gallagher and Appenzeller, 1999). On the other hand, as West and West (2013, page 210) once pointed out, "…science does not wait for definitions, it continues forward in exploring phenomena with or without a clear understanding, confident that such understanding will eventually emerge." Saint Thomas Aquinas said: 'What, then, is time? If no one asks me, I know what it is. If I wish to explain it to him who asks me, I do not know.' Now, for me, 'What, then, is city/fractal/science? If no one asks me, I know what it is. If I wish to explain it to him who asks me, I do not know.' Even so, as Potter Stewart, the well-known former judge of the United States, said, "I know it when I see it." (Cited from Arbesman, 2012). I know if it is a city when I see a city, I know if it is a fractal when I see a fractal, and I know if it is scientific research when I see a research result.

## 5. Conclusions

Fractal geometry provides us a new mathematical framework of describing urban morphology. To characterize urban form and explain urban growth, we need various fractal dimensions. Fractal dimensions can be defined by generalized entropy and correlation function. To understand the essence of fractal dimension, we must learn about entropy and correlation function. On the one hand, fractal dimension is a characteristic value of entropy, and on the other, fractal dimension is a scaling exponent of correlation function. Where entropy is concerned, fractal dimension indicates uniformity, and inequality degree and uniformity degree represents two different sides of the same coin. In this sense, fractal dimension suggests difference and diversity. Where correlation is concerned, fractal dimension implies complexity degree of dynamical systems. Moreover, to understand fractal dimension concept, we should know the notions of topological dimension and Euclidean dimension of embedding space in which fractal cities are defined. Fractal theory can be employed to make spatial analysis for the scale-free aspects of urban morphology. To research urban growth, we can employ sigmoid functions to model fractal dimension growth curves of urban form based on time series of fractal dimension. Thus, we have to know the upper limit and lower limit of fractal dimension values. The lower limit of fractal dimension relates to the topological dimension of fractal sets, while the upper limit depends on the embedding space dimension.

The main points of this paper can be summarized as follows. *First, fractal geometry is powerful tool of scale-free analysis, and urban morphology is typical scale-free geographical phenomenon. Therefore, fractal theory can be naturally applied to urban studies.* Cities are not true fractals, but they can be treated as random pre-fractals, which bear fractal properties within certain scaling ranges. If urban form had characteristic scales, we would be able to calculate urban area and urban perimeters. Thus urban form can be described with the methods from traditional advanced mathematics. Unfortunately, urban form has no characteristic scales, it belong to scale-free distributions. A great many studies show that urban form follow power laws indicative of fractal nature. In this case, it is an advisable selection to employ fractal geometry to describe urban morphology and make scaling analysis of urban patterns and dynamic processes. *Second, the most proper dimension of embedding space for city fractals is 2 dimension rather than 3 dimension. The upper limit of fractal dimension of urban form should not exceed the embedding dimension.* A city fractal can be defined in a 2-dimensional space, and it can also be defined in a 3-dimensional space. It is better to define city fractals in a 2-dimensional space. On the one hand, fractal dimension is used to replace urban area, which cannot be objectively measured due to scale-free distribution of cities. Urban area is a measure defined in 2-dimensional space. Therefore, city fractals can be defined in 2-dimensional space so that fractal dimension can be employed to well replace urban area. On the other hand, the criterion of scientific method is to reduce dimensions rather than increase dimensions. Moreover, more available datasets of cities are based on 2-dimensional space. It is simpler and more effective to analyze a city fractal through 2-dimensional space. *Third, the topological dimension of urban form is 0 dimension rather than 1 dimension. The lower limit of fractal dimension is equal to or greater than the topological dimension.* For modeling fractal dimension growth curves of urban form, it is significant to identify the lower limit of fractal dimension. In theory, urban form can be reduced to separated spaces, so the topological dimension of city fractals is $d_T=0$. The lower limit of fractal dimension of urban form is $D_{min}=0$. The topological dimension of urban boundary is 1, but the most important city fractals are based on urban area instead of urban boundary. In practice, the lower limit of fractal dimension of urban form can also be treated as $D_{min}=1$ especially when the sample path is short. Based on the constant study area and fixed largest box, the lower limit of fractal dimension of urban form should be taken as $D_{min}=0$. Based on the variable study area and unfixed largest box, the lower limit of fractal dimension of urban form should be taken as $D_{min}=1$. Based on the constant study area, fixed largest box and long sample path (time span is very large), the fractal dimension values of urban form is sometimes $D<1$. How to take the $D_{min}$ value, it depends on the concrete situation.

## Acknowledgements

This research was sponsored by the National Natural Science Foundations of China (Grant No. 41671167). The support is gratefully acknowledged.

Chinese]

## Appendix—Derivation of the relationships between fractal dimension and standard error

The relationships between fractal dimension and standard error can be derived with the help of knowledge of linear algebra and statistics. The following regression coefficient formula and basic statistics such as correlation coefficient $R$, and $F$ and $t$ statistics can be seen in many statistical analysis textbooks, and will not be explained in detail. The fractal model can be expressed as a power function as below:

$$N(r) = Kr^{-D}, \tag{A1}$$

where $r$ denotes the spatial measurement scale such as linear size of box, $N(r)$ is the corresponding spatial measurement with linear size $r$ such as the number of nonempty boxes, $K$ is the proportionality coefficient, and $D$ is the fractal dimension. The natural logarithm of both sides of the equation (A1) is

$$\ln N(r) = \ln K - D \ln r. \tag{A2}$$

Then, the relationships between fractal dimension and its standard error can be deduced in two cases.

(1) Free intercept regression. Suppose that $K>0$ but $K \neq 1$. Thus we have free intercept regression. In this case, the degree of freedom is $v=n-2$. For simplicity, equation (A2) is expressed as a univariate linear regression equation as follows

$$y = a + bx, \tag{A3}$$

in which $x=\ln(r)$ refers to the independent variable and $y=\ln N(r)$ to the dependent variable. As for the parameters, $a=\ln K$ denotes the intercept, and $b=-D$ is the slope. The slope value is termed regression coefficient in linear regression analysis. By the idea of least error sum of squares, we can construct a normal equation system. Suppose the time of measurements is $n$, and the measurement sequence is numbered as $i=1,2,\ldots, n$. Then by using Cramer rule, we can derived the formula of the regression coefficient

$$b = \frac{\sum_{i=1}^{n}(x_i - \overline{x})(y_i - \overline{y})}{\sum_{i=1}^{n}(x_i - \overline{x})^2}, \tag{A4}$$

where $\overline{x}$ , $\overline{y}$ represent arithmetic means of the independent variable $x$ and dependent variable $y$, respectively. The multiple correlation coefficient is defined as

$$R^2 = \frac{(\sum_{i=1}^{n}(x_i - \overline{x})(y_i - \overline{y}))^2}{\sum_{i=1}^{n}(x_i - \overline{x})^2 \sum_{i=1}^{n}(y_i - \overline{y})^2} = \frac{\sum_{i=1}^{n}(\hat{y}_i - \overline{y})^2}{\sum_{i=1}^{n}(y_i - \overline{y})^2} = 1 - \frac{\sum_{i=1}^{n}(y_i - \hat{y}_i)^2}{\sum_{i=1}^{n}(y_i - \overline{y})^2}. \tag{A5}$$

Substituting equation (A5) into equation (A4) yields

$$b = \frac{\sum_{i=1}^{n}(x_i - \overline{x})(y_i - \overline{y})}{\sqrt{\sum_{i=1}^{n}(x_i - \overline{x})^2 \sum_{i=1}^{n}(y_i - \overline{y})^2}} \sqrt{\frac{\sum_{i=1}^{n}(y_i - \overline{y})^2}{\sum_{i=1}^{n}(x_i - \overline{x})^2}} = R\sqrt{\frac{\sum_{i=1}^{n}(y_i - \overline{y})^2}{\sum_{i=1}^{n}(x_i - \overline{x})^2}}. \tag{A6}$$

In fact, equation (A4) and equation (A5) can be found in many textbooks of multiple statistical analyses. The $F$ statistic is defined as

$$F = \frac{\sum_{i=1}^{n}(\hat{y}_i - \overline{y})^2}{\frac{1}{n-2}\sum_{i=1}^{n}(y_i - \hat{y}_i)^2}. \tag{A7}$$

Substituting equation (A5) into equation (A7) yields

$$F = \frac{(n-2)\sum_{i=1}^{n}(\hat{y}_i - \overline{y})^2 / \sum_{i=1}^{n}(y_i - \overline{y})^2}{\sum_{i=1}^{n}(y_i - \hat{y}_i)^2 / \sum_{i=1}^{n}(y_i - \overline{y})^2} = \frac{(n-2)R^2}{(1-R^2)}. \tag{A8}$$

The $t$ statistic is defined as

$$t = \frac{b}{s}, \tag{A9}$$

where $s$ refers to the standard error of the regression coefficient based on sample. In contrast, the standard error $\delta$ is based on population. The formula is

$$s=\sqrt{\frac{\frac{1}{n-2}\sum_{i=1}^{n}(y_i-\hat{y}_i)^2}{\sum_{i=1}^{n}(x_i-\overline{x})^2}}\,. \qquad \text{(A10)}$$

Equations (A7), (A8), (A9), and (A10) can be found in many statistics textbooks. Substituting equations (A5), (A6) and (A10) into equation (A9) yields

$$t=R\sqrt{\frac{\sum_{i=1}^{n}(y_i-\overline{y})^2}{\sum_{i=1}^{n}(x_i-\overline{x})^2}}\Bigg/\sqrt{\frac{\frac{1}{n-2}\sum_{i=1}^{n}(y_i-\hat{y}_i)^2}{\sum_{i=1}^{n}(x_i-\overline{x})^2}}=\sqrt{\frac{\sum_{i=1}^{n}(\hat{y}_i-\overline{y})^2}{\frac{1}{n-2}\sum_{i=1}^{n}(y_i-\hat{y}_i)^2}}=\sqrt{F}\,, \qquad \text{(A11)}$$

Combining equations (A8) and (A11), we have

$$F=t^2=\frac{(n-2)R^2}{(1-R^2)}\,. \qquad \text{(A12)}$$

Substituting equations (A9) into equation (A12) yields

$$\frac{|b|}{s}=\sqrt{\frac{(n-2)R^2}{(1-R^2)}}\,. \qquad \text{(A13)}$$

Considering the definition of slope given above, $b$=-$D$, we have

$$s=|b|\sqrt{\frac{1-R^2}{(n-2)R^2}}=D\sqrt{\frac{1/R^2-1}{n-2}}\,. \qquad \text{(A14)}$$

Thus, based on the 95% level of confidence, corresponding to the significance level of 0.05, the margin of error of the fractal dimension value can be expressed as

$$D\pm \mathrm{tinv}(\alpha,n-2)\cdot D\sqrt{\frac{1/R^2-1}{n-2}}\approx D(1\pm 2\sqrt{\frac{1/R^2-1}{n-2}})=D\pm 2s\,. \qquad \text{(A15)}$$

where tinv is the MS Excel function for threshold value investigation of the $t$ statistic, $\alpha$ denotes significance level. The grammar the function “tinv” is “tinv($\alpha$, $v$)”, and here the level of significance should be $\alpha$=0.05.

(2) Free intercept regression. If $K$=1, then $a$=ln$K$=0. Thus we have fixed intercept regression. In this case, the degree of freedom is $v$=$n$-1. Then, the Pearson coefficient is replaced by the cosine formula, that is

$$R^2 = \frac{(\sum_{i=1}^{n} x_i y_i)^2}{\sum_{i=1}^{n} x_i^2 \sum_{i=1}^{n} y_i^2}. \tag{A16}$$

Using the similar method, we can derive the relation between fractal dimension and standard error based on fixed intercept regression as below

$$s = D\sqrt{\frac{1/R^2 - 1}{n-1}}. \tag{A17}$$

In order to save space, the detailed derivation of equation (A17) is omitted. Readers can derive it by analogy with the process of deriving equation (A14). Based on the 95% level of confidence, the margin of error of the fractal dimension value is

$$D \pm \text{tinv}(\alpha, n-1) \cdot D\sqrt{\frac{1/R^2 - 1}{n-1}} \approx D(1 \pm 2\sqrt{\frac{1/R^2 - 1}{n-1}}) = D \pm 2s. \tag{A18}$$

This means that, based on the 95% confidence level, the error margin of the fractal dimension is approximately the fractal dimension value plus or minus twice the standard error. Note that the sample standard error $s$ here is used instead of the population standard error $\delta$ in the text. The population standard error $\delta$ is mainly for theoretical derivation, while the sample standard error $s$ is principally for empirical analyses.